# Distance determination method of dust particles using Rosetta OSIRIS NAC and WAC data

E. Drolshagen[1]\*, T. Ott[1]\*, D. Koschny[2, 4], C. Güttler[3], C. Tubiana[3], J. Agarwal[3], H. Sierks[3], C. Barbieri[5], P. L. Lamy[6], R. Rodrigo[7,8], H. Rickman[9,10], M. F. A'Hearn[3,11,12], M. A. Barucci[13], J.-L. Bertaux[14], I. Bertini[15], G. Cremonese[16], V. Da Deppo[17], B. Davidsson[9], S. Debei[18], M. De Cecco[19], J. Deller[3], C. Feller[13], S. Fonasier[13], M. Fulle[19], A. Gicquel[3], O. Groussin[20], P. J. Gutiérrez[21], M. Hofmann[3], S. F. Hviid[3,22], W.-H. Ip[23,24], L. Jorda[20], H. U. Keller[22,25], J. Knollenberg[22], J. R. Kramm[3], E. Kührt[22], M. Küppers[26], L. M. Lara[21], M. Lazzarin[5], J. J. Lopez Moreno[21], F. Marzari[5], G. Naletto[15,17,27], N. Oklay[3], X. Shi[3], N. Thomas[28,29], and B. Poppe[1]

\*These authors contributed equally to this work

The ESA Rosetta spacecraft has been tracking its target, the Jupiter-family comet 67P/Churyumov-Gerasimenko, in close vicinity for over two years. It hosts the OSIRIS instruments: the Optical, Spectroscopic, and Infrared Remote Imaging System composed of two cameras, see e.g. Keller et al. (2007). In some imaging sequences dedicated to observe dust particles in the comet's coma, the two cameras took images at the same time.

The aim of this work is to use these simultaneous double camera observations to calculate the dust particles' distance to the spacecraft. As the two cameras are mounted on the spacecraft with an offset of 70 cm, the distance of particles observed by both cameras can be determined by a shift of the particles' apparent trails on the images. This paper presents first results of the ongoing work, introducing the distance determination method for the OSIRIS instrument and the analysis of an example particle. We note that this method works for particles in the range of about 500 m – 6000 m from the spacecraft.

## 1 Introduction

The ESA Rosetta spacecraft was launched in 2004 and arrived at its main target, the Jupiter-family comet 67P/Churyumov-Gerasimenko (hereafter 67P), in 2014. The scientific camera system onboard Rosetta is the Optical, Spectroscopic, and Infrared Remote Imaging System, OSIRIS (Keller et al. 2007). It consists of two cameras, a Narrow Angle Camera (NAC), with a field of view (FOV) of 2.20° x 2.22°, and a Wide Angle Camera (WAC), which has a FOV of 11.35° x 12.11°. Both cameras are equipped with a 2048 x 2048 pixel CCD detector (Keller et al. 2007).

The combination of the NAC and the WAC serves the following scientific objectives of the camera system on Rosetta. On the one hand, the NAC, a camera with high spatial resolution, enables an early detection of the nucleus. It makes it possible to examine the structure of the nucleus as well as its rotation from distance, to study the surface mineralogy, and to investigate the process of dust ejection. On the other hand, the WAC, a camera with lower spatial resolution has a larger field of view. This offers information of the dynamics of dust and gas around the nucleus. Furthermore, the WAC allows a view of the nucleus in its entirety. Overall, the WAC enables observations of the whole nucleus over a long period when Rosetta is close to the nucleus and the NAC studies it in a more detailed way. This is explained in more detail in Keller et al. (2007).

Studying the diverse processes occurring on 67P is one of the aims of the Rosetta mission. One goal is the investigation of the size distribution of dust aggregates on the nucleus as well as in its close vicinity. This knowledge will enhance the understanding of the evolution of comets. It is

Esther Drolshagen: Eshter.Drolshagen@uni-oldenburg.de
Theresa Ott: Theresa.Ott@uni-oldenburg.de

possible to measure the dust flux in the coma utilizing optical images as well as dust detectors (Fulle et al. 2016, Rotundi et al. 2015). Rosetta has been monitoring 67P from August 2014 to September 2016 from close vicinity. That enables the detailed study of the dust size distribution of the comet. Furthermore, it can be investigated if and how the distribution evolves in time, and the connection with the nucleus seasons. With detections of dust particles, collected between 3.6 and 3.4 AU inbound, it was possible to derive a dust size distribution for a large mass bin range ($10^{-10} - 10^{-2}$ kg) (Fulle et al. 2016, Rotundi et al. 2015). The analysis was based on data from the GIADA instrument, see e.g. Della Corte et al. (2014), and the OSIRIS cameras.

One of the first steps of OSIRIS particle mass determination is defining the distance of the dust particles to the cameras. This can be achieved with different methods. One of them is using the parallax effect as shown by Fulle et al. (2016). They assumed that the dust motion is mostly radial from the nucleus, so any other observed dust motion is due to spacecraft motion. This provides the necessary information to conduct parallax determinations utilizing the spacecraft velocity and the exposure times. With this technique it is possible to analyze a large range of particles with different distances (Fulle et al. 2016). Davidsson et al. (2015) determined the orbits of four particles by utilizing the parallax. They used WAC image sequences with several consecutively taken images and determined the particle movement relative to the background stars.

In this work we present a method that makes use of both cameras. The NAC and WAC are placed about 0.7 m apart from each other on the spacecraft. Exploiting this set up it is possible to derive the distances of particles to the cameras using the parallax effect with a high accuracy and very few assumptions.

In some imaging sequences dedicated to dust-particle tracking the two cameras took images at the same time. From these double camera observations it is possible to determine the geometry of the simultaneously detected dust trails. In the following section the data set and the image processing routines will be presented. We analyzed particles that show a parallax shift between the trails seen by NAC and by WAC. To determine the distance, the start and end positions of the particle trails have to be found. In Section 3 a constructed problem will show that, in theory, these are the positions where the intensity of the profile of the particle has fallen to half its maximum.

Vereš et al. (2012) published a similar method about the detection and analyses of asteroid data. The asteroids appear as trails in the analyzed ground-based images. For the detection of the trails of these asteroids they used an analytic function assuming a Gaussian point spread function that moves with a constant velocity. Showing an increased accuracy of their method compared with assuming a 2-dimensional Gaussian to fit the profile of the particle. The resulting signal is derived as an integral of Gaussians. Although, Vereš et al. analyzed asteroid trails with mean lengths of 15 pixels the fitting function presented herein is very similar and derived the same way. The trails analyzed in the OSIRIS images are much longer and the in-situ data is free of affects like seeing. Moreover, the NAC and WAC data is calibrated for flux (Tubiana et al. (2015)) which is important for trail shaped signals as shown by Fraser et al. (2016).

The determination of the particle distance to the spacecraft, based on the parallax effect, will be further explained in Section 4. In Section 5.1 this work presents the distance determination of one particle as part of our ongoing work.

## 2 The data

The so-called GRAIN_COLOR sequences were taken for the purpose of double camera observations. In these sequences the two scientific cameras onboard Rosetta, both pointing in almost the same direction, took images at nearly the same time, differing only by a small fraction of the entire exposure time. In these sequences the WAC always took an image with a long exposure time (first 64 s and later 90 s) while the NAC took three or four images consecutively in different color filters, each with an exposure time of 12.5 s. The utilized filters are shown in Table 1. All WAC data was collected by using the green filter (F21). The NAC utilizes four different filters: F41, (F32 if the camera took four images), F22, and F24 (see Table 1 for details about the filters). The filter F32 uses the same orange filter as F22 but in combination with a different focus and was added in an attempt to utilize the different focus distances of the Near Focus Plate (used by F32) and the Far Focus Plate (used by F22) to determine the distance via defocusing. Between the individual NAC images there are breaks of the exposures of 11 s in which the camera changes the filter. This results in a time range for the NAC image sets of 59.5 s, 83 s, respectively. The mid-point of the exposure of the WAC image and the mid-point of the NAC sequence were synchronized for the observations. In Fig. 1 one image sequence is presented. It shows the four corresponding images of the first image set, taken on 24 July 2015 at about 15:21 UTC. a) - c) show the consecutively taken NAC images and d) shows the WAC image taken simultaneously. The data have been radiometrically calibrated and geometrically corrected using the OSIRIS calibration pipeline (see Tubiana et al. (2015)). The distortion of the images due to the optics is less than 1.5 % for the NAC over the entire FOV (Dohlen et al. 1996).

The WAC has a FOV of about 12° and the NAC of only about 2°. According to the FOV ratios (elevation: 0.194, azimuth: 0.183), the WAC images have to be cropped, compare the red box in Fig. 1 d).

To stack the consecutive NAC images, the frames have to be aligned using stars visible in the image. This is due to the fact that the pointing position of the camera is not exactly the same for each exposure of a set. Not only are there small calculated changes but also random motions that are caused by resonances of the solar arrays or by particles hitting the spacecraft and its solar arrays. The alignment yields

R3

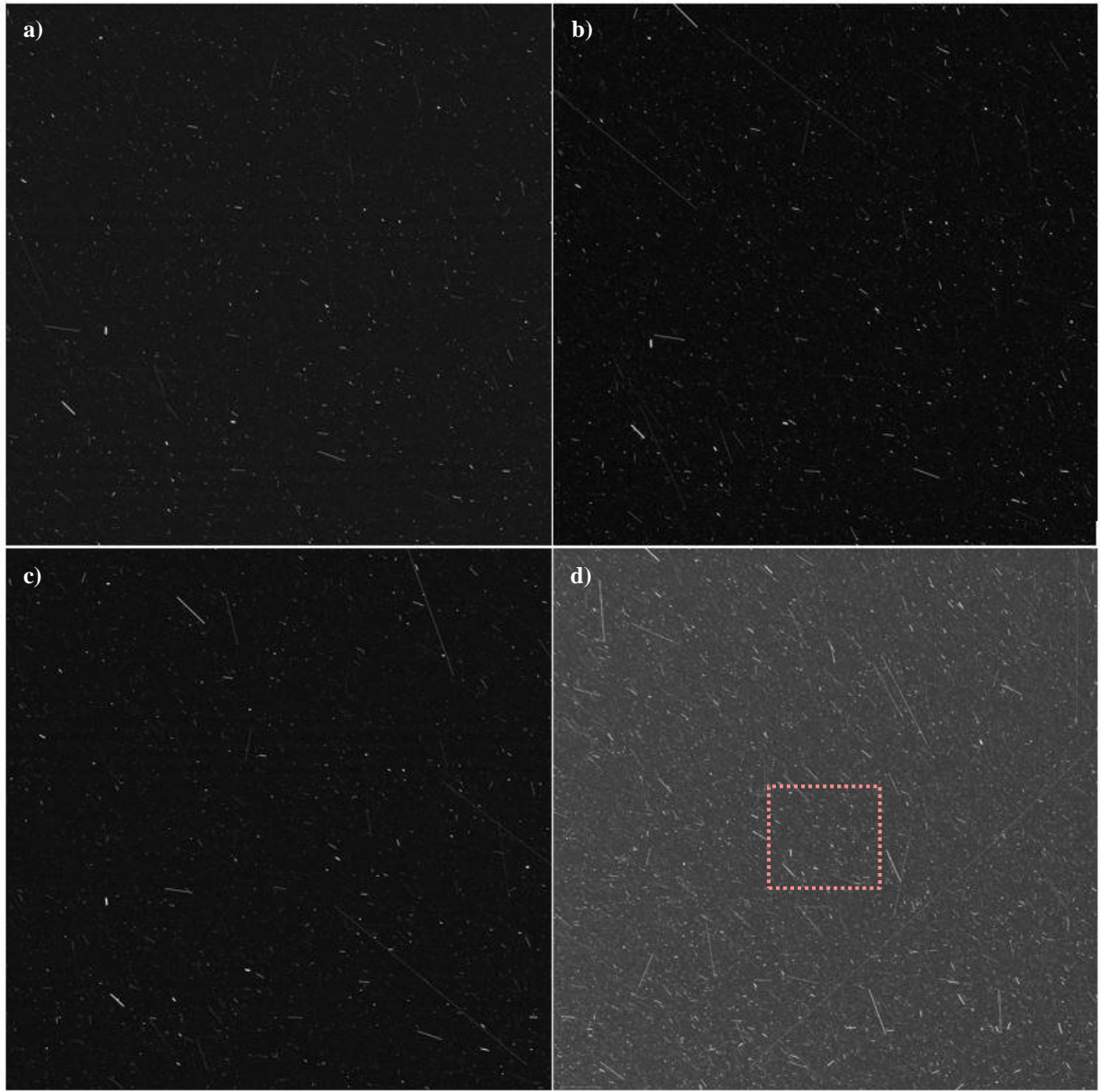

**Figure 1** One image set taken on 24 July 2015 at 15:22:13 UTC. a) - c) show the NAC images taken consecutively and d) the simultaneously recorded WAC image. The original filenames are a) NAC_2015-07-24T15.20.55.765Z_ID10_1397549000_F41, b) NAC_2015-07-24T15.21.19.869Z_ID10_1397549001_F22, c) NAC_2015-07-24T15.21.43.739Z_ID10_1397549002_F24, and d) WAC_2015-07-24T15.20.54.420Z_ID10_1397549100_F21. The red dotted box shows the area that corresponds to the FOV of the NAC.

Esther Drolshagen: Eshter.Drolshagen@uni-oldenburg.de
Theresa Ott: Theresa.Ott@uni-oldenburg.de

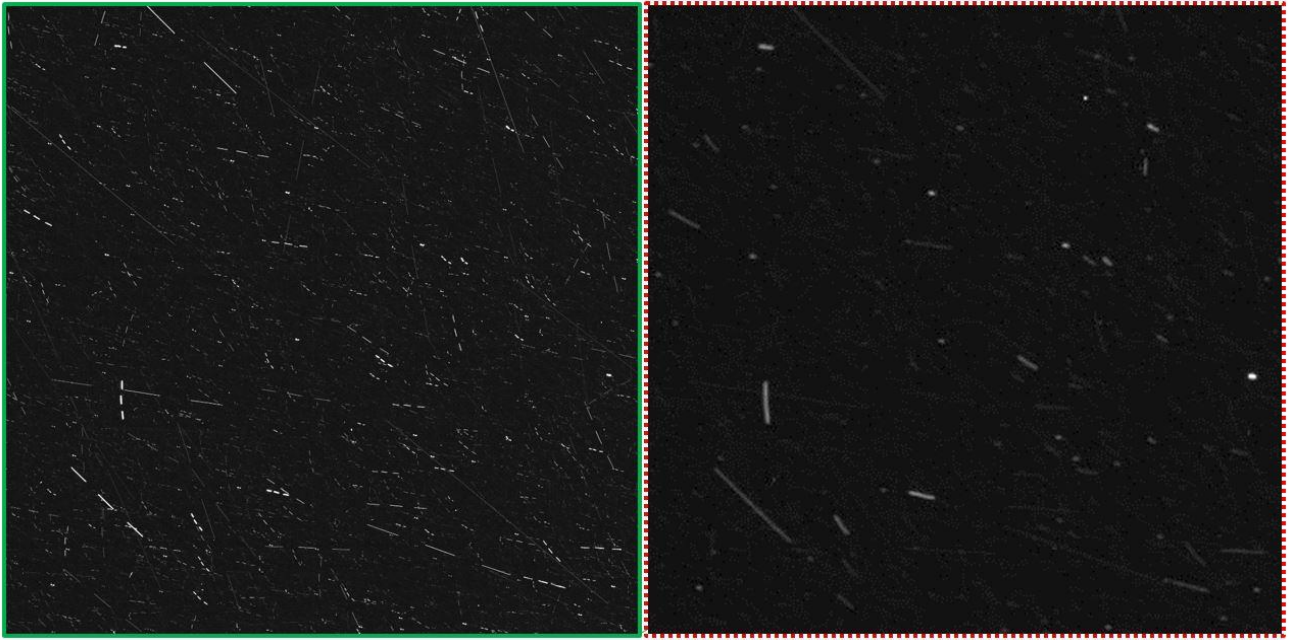

**Figure 2** The aligned, and (a) stacked NAC and (b) cropped WAC image. The processed data is the same as shown in Fig. 1.

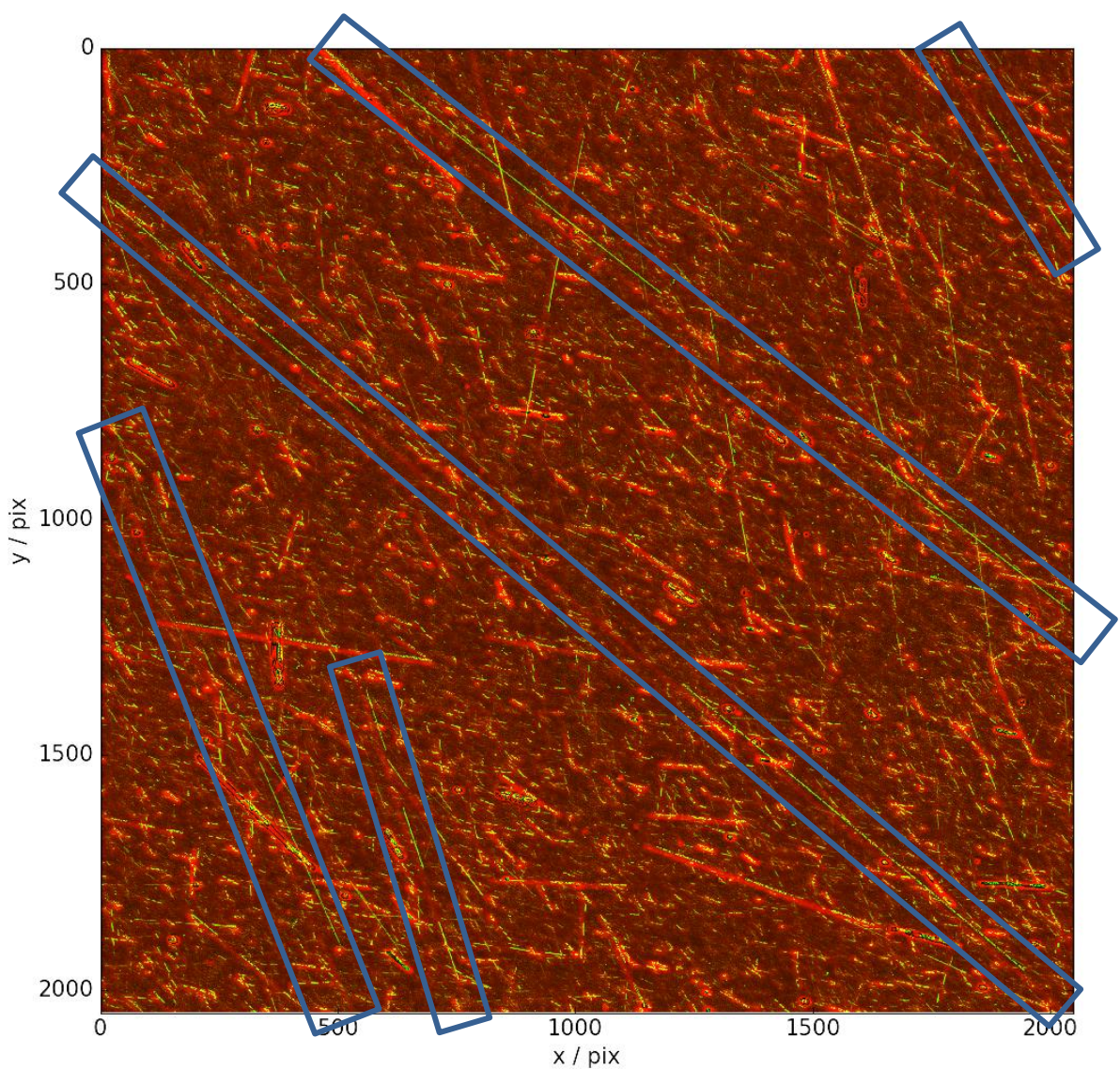

**Figure 3** The aligned NAC and WAC data as a red-green image. The stacked NAC image is displayed in green (particles visible as up to three separate trails) and the cropped WAC image in red (particles visible as one solid trail). The processed data are the same as presented in Fig. 1. In this image set there are five dust particles found that have a shift of more than six pixels between the NAC and WAC trail, highlighted with blue boxes.

results with an accuracy on a subpixel scale needed for the determination of the positions of the included particle trails.

The cameras are both pointed at one position during the exposure time. Consequently, in long exposures stars are untrailed, point-like objects on a CCD. For the alignment the software PixInsight and its process DynamicAlignment was used (Pleiades Astrophoto S.L. 1.8, Valencia, Spain). It is a semi-automatic image registration system aligning images by using hand-picked reference stars. The first NAC image will always be used as the reference image to which the others are aligned, compare Fig. 1 a). The same six stars were chosen manually for an image set. This was done for all images of the sequence. Afterwards, the aligned NAC images can be stacked utilizing the maximum pixel values.

The results are presented in Fig. 2. Due to the fact that the analyzed NAC images are stacked combinations of three or four images, the NAC dust trails show as three to four aligned tracks. Additionally, it has to be kept in mind that the FOV of the WAC is much larger than the one of the NAC. That is the reason why only the dotted rectangle in the image Fig. 1 d) corresponds to a) – c). Compare the stacked NAC images in Fig. 2 a) with the cropped WAC image in Fig. 2 b).

Afterwards, the stacked NAC image and the WAC image were merged into different color channels of the same image. The result is shown in Fig. 3, where the stacked NAC image is shown in green and the cropped WAC image in red. In this set there are five particles that display a shift between the NAC and WAC trail of more than six pixels. They are highlighted with solid boxes. It can be seen that most analyzed particles have produced only weak signals.

**Table 1** The filter numbers and names, which the NAC and WAC use, whereas all data the WAC collected is taken utilizing F21, the "Green Empty" filter (Keller et al. 2007). The second orange filter (F32) was added to the sequences in an attempt to use the different focus distances of the Far Focus Plate (FFP) and the Near Focus Plate (NFP) to determine the distance via defocusing.

| Camera | Filter Number | Filter Name | Central Wavelength / nm |
|---|---|---|---|
| WAC | 21 | Green_Empty | 535.7 |
| NAC | 41 | Near-IR_FFP-IR | 882.1 |
| NAC | 22 | FFP-Vis_Orange | 649.2 |
| NAC | 24 | FFP-Vis_Blue | 480.7 |
| NAC | 32 | NFP-Vis_Orange | 649.2 |

## 3 Theoretical signal – moving 2-dimensional Gaussian

This section explains the formation of the intensity profile of dust particle signals in space. To do this, a constructed theoretical particle will be analyzed. In one moment of exposure, point sources (like stars) have a Gaussian-shaped signal on the CCD of the recording cameras. However, a dust particle is not fixed and moves through the FOV, relative to the stars. The particle would be visible as an elongated trail. This is similar to the approach Vereš et al. (2012) took in their paper.

### 3.1 The signal

When the camera starts the exposure, the dust particle is a Gaussian-shaped point on the CCD, similar to the signal of stars. The general rotationally symmetric 2-dimensional Gaussian in two dimensions is given by Eq. (1). There, $A$ is the amplitude, $x_0, y_0$ are the center coordinates, and $\sigma_x, \sigma_y$ are the x respectively y spreads of the 2-dimensional Gaussian. Assuming that the bell curve is rotationally symmetric yields $\sigma_x = \sigma_y =: \sigma$.

$$f(x,y) = A \cdot \exp\left(-\left(\frac{(x-x_0)^2}{2\cdot\sigma^2}\right) - \left(\frac{(y-y_0)^2}{2\cdot\sigma^2}\right)\right) \quad (1)$$

To simplify the problem we assume that the movement of the particle is along the x-axis and at each position a 2-dimensional Gaussian is recorded. Accordingly, $x_0$ changes over time and $y_0$ remains unchanged. Furthermore, the trail will be created in a way that at the beginning of the exposure time the particles center is placed at exactly $(0,0)$. Then, the Gaussian that is recorded in the first moment of the exposure can be expressed with Eq. (2). The consecutive ones can be described by Eq. (3). In Eq. (3), $s(t) = k \cdot t$ is the distance in x-direction of the particle center at the time $t$ from the starting position. The distance increases linearly, since we assume a constant motion of the particle, which is represented by the constant factor $k$. Illustrating the problem, the Gaussian described in Eq. (2) is plotted in Fig. 4 a) in a two-dimensional map and in b) in three dimensions. $A = 1$ and $\sigma = 1$ were used for the plots.

$$f(x,y,t=0) = A \cdot \exp\left(-\left(\frac{x^2}{2\cdot\sigma^2}\right) - \left(\frac{y^2}{2\cdot\sigma^2}\right)\right) \quad (2)$$

$$f(x,y,t=1,\ldots,N) = A \cdot \exp\left(-\left(\frac{(x-s(t))^2}{2\cdot\sigma^2}\right) - \left(\frac{y^2}{2\cdot\sigma^2}\right)\right) \quad (3)$$

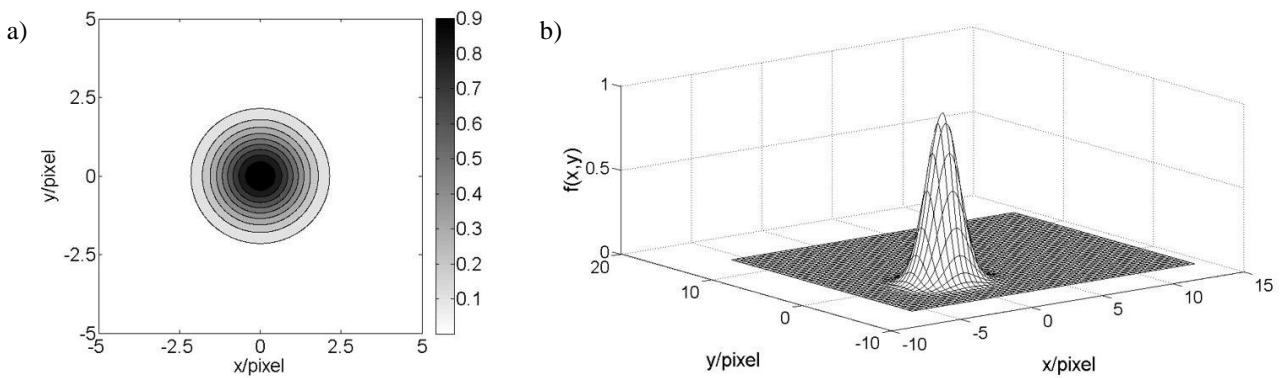

**Figure 4** The 2-dimensional Gaussian described in Eq. (2) with $A = 1$ and $\sigma = 1$; a) 2D map; b) 3D plot.

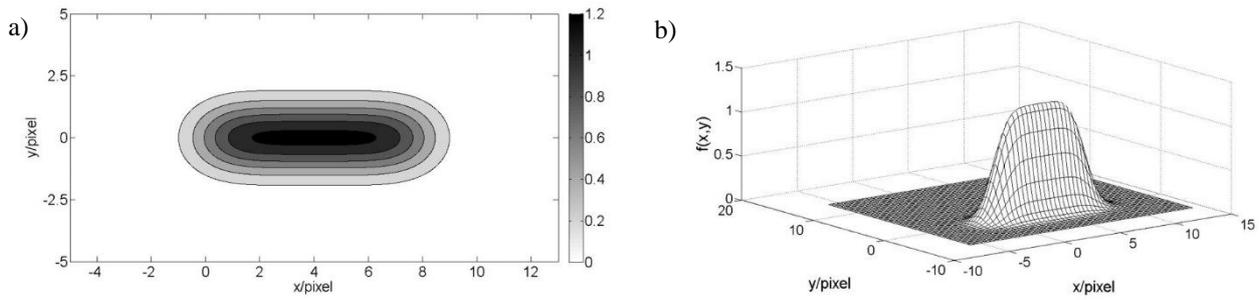

**Figure 5** The resulting function that describes a dust particles trail, see Eq. (10) with $A = 1$, $\sigma = 1$, and $N = 8$; a) 2D plot; b) 3D plot.

The complete moving particles signal recorded by the CCD can be expressed as the sum over all Gaussians with $s(t) = k \cdot t$, where $k$ is the time difference between the exposures, and a repetition of the exposure of $N$:

$$f(x,y) = \sum_{t=0}^{N} A \cdot \exp\left(-\left(\frac{(x-k \cdot t)^2}{2 \cdot \sigma^2}\right) - \left(\frac{y^2}{2 \cdot \sigma^2}\right)\right) \quad (4)$$

Between the single bell curves the steps are actually infinitesimal small, consequently, the sum has to be converted into an integral:

$$f(x,y) = \int_0^N A \cdot \exp\left(-\left(\frac{(x-t)^2}{2 \cdot \sigma^2}\right) - \left(\frac{y^2}{2 \cdot \sigma^2}\right)\right) dt \quad (5)$$

Only the first part of the argument of the exponential function depends on $t$, leading to Eq. (6). Hence, the integral that states in Eq. (7) has to be solved.

$$f(x,y) = A \cdot \exp\left(-\frac{y^2}{2 \cdot \sigma^2}\right) \int_0^N \exp\left(-\left(\frac{(x-t)^2}{2 \cdot \sigma^2}\right)\right) dt \quad (6)$$

$$g(x) = \int_0^N \exp\left(-\left(\frac{(x-t)^2}{2 \cdot \sigma^2}\right)\right) dt \quad (7)$$

To do so, first a substitution has to be conducted and then the integral can be split at zero into two integrals. The result with the reversed interval of integration of the first term is presented in Eq. (8).

$$g(x) = -\sigma \cdot \sqrt{2} \cdot \left[\int_0^{\frac{x-N}{\sigma \cdot \sqrt{2}}} \exp(-\varepsilon^2) \, d\varepsilon - \int_0^{\frac{x}{\sigma \cdot \sqrt{2}}} \exp(-\varepsilon^2) \, d\varepsilon\right] \quad (8)$$

To solve that expression the error function ($erf(\kappa)$) is needed. (For further details about the error function see e.g. Andrews (1998).)

$$\text{erf}(\kappa) = \frac{2}{\sqrt{\pi}} \int_0^\kappa \exp(-\tau^2) \, d\tau. \quad (9)$$

Consequently, $f(x,y)$ can be expressed as:

$$f(x,y) = -A \cdot \sigma \cdot \sqrt{\frac{\pi}{2}} \cdot \exp\left(-\frac{y^2}{2 \cdot \sigma^2}\right) \cdot \left[\text{erf}\left(\frac{x-N}{\sigma \cdot \sqrt{2}}\right) - \text{erf}\left(\frac{x}{\sigma \cdot \sqrt{2}}\right)\right] \quad (10)$$

This solution is presented in Fig. 5 a) in two dimensions and in b) in three dimensions, again with $A = 1$, $\sigma = 1$, and $N = 8$.

### 3.2 The intensity profile

To determine the beginning and end points of the observed particles, the profile of the theoretical signal has to be further investigated. Since the movement is only along the x-axis, the intensity profile along that axis of the resulting signal looks like the graph in Fig. 6. The plotted function is the same as in Fig. 5. We will show that the beginning and end points of the theoretical particle lay at the half maximum point of the profile through the particle.

#### 3.2.1 The edges

The 2-dimensional Gaussians are positioned with their centers along the x-axis. Consequently, the first part of Eq. (10) can be neglected because it only depends on y. This yields Eq. (11). In the code the trail along the particle is set as the x-axis, transforming the problem to the same geometry. The intensity profile is shown in Fig. 6. A closer look at Fig. 6 shows that the maximum of the intensity profile function is part of the plateau in the center of the profile.

Next, the start and end points of the trail are investigated using the intensity profile values, they are drawn in as 'x' and '+' in Fig. 6. These points seem to have an intensity value half as high as the maximum of the function. To investigate this assertion, we are looking at the relation of the maximum point in the center of the profile function at $x_M = N/2$, and the theoretical start point at $x_{start} = 0$. The equation to solve is stated in Eq. (14) left. Additionally, the property of the error function stated in Eq. (12) and Eq. (13) (Andrews 1998) can be used to simplify the error function parts of Eq. (11), see Eq. (14). Consequently, $f(0)/f(N/2) \approx 1/2$ for trails that are much larger than their Full Width Half Maximum (FWHM) and hence their $N/\sigma$ is large enough to allow the approximation.

$$f(x) = -A \cdot \sigma \cdot \sqrt{\frac{\pi}{2}} \cdot \left[erf\left(\frac{x-N}{\sigma \cdot \sqrt{2}}\right) - erf\left(\frac{x}{\sigma \cdot \sqrt{2}}\right)\right] \quad (11)$$

$$erf(-x) = -erf(x) \quad (12)$$

$$erf(0) = 0 \quad (13)$$

$$\frac{f(0)}{f\left(\frac{N}{2}\right)} = \frac{1}{2} \cdot \frac{-\text{erf}\left(\frac{N}{\sigma \cdot \sqrt{2}}\right)}{-\text{erf}\left(\frac{\frac{N}{2}}{\sigma \cdot \sqrt{2}}\right)} \approx \frac{1}{2} \quad (14)$$

We can check this assumption by investigating the FWHM and the length of the particles. In Magrin et al. (2015) the FWHM of the signal on the CCD of the star Vega is measured for different filters for the two OSIRIS cameras. The values are at around 2 pixels, see their fig. 2. In every moment, the signal of the dust particle is comparable to one of a star. Since the general FWHM is dependent on $\sigma$ as given by Eq. (15), $\sigma$ can be calculated with the value for the FWHM. We used an approximation of 2 pixels, utilizing the values from Magrin et al. (2015).

$$FWHM = 2 \cdot \sigma \cdot \sqrt{2 \cdot \ln(2)} \approx 2 \quad (15)$$

$$\sigma \approx (2 \cdot \ln(2))^{-\frac{1}{2}} \approx 1 \quad (16)$$

A first look at the images with particles to be determined show a variety of particle lengths. We expect to find most particles to be under 500 pixel long with a mean around 350 pixels. The data is relatively noisy due to the high number of particles in the images. We expect low SNR for most particles. This hinders the detection. Consequently, we analyze particles with large N and a $\sigma \approx 1$ allowing the approximation in Eq. (14) to be used to find the start position of the profile. Since the profile of the particle is symmetric around $x_M$, this also works for $x_{end} = N = 8$. Accordingly, the start and end positions of the particle correspond to the profile points where the intensity has fallen to half its maximum value.

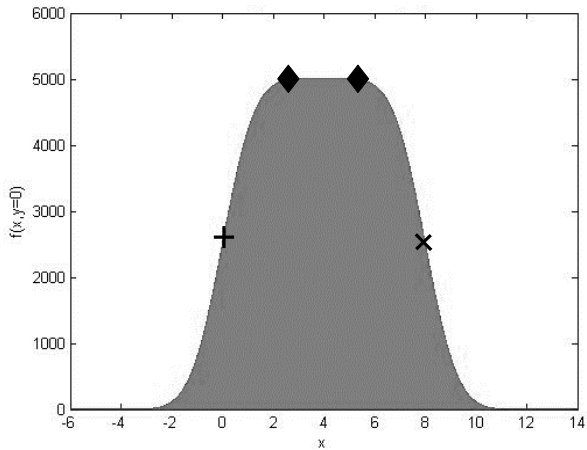

**Figure 6** The intensity profile of the resulting signal along the x-axis. The function is the same as in Fig. 5 with $A = 1, \sigma = 1$, and $N = 8$. The maximum is centered at the profile and the beginning and end of the plateau are marked with diamonds. The start and end of the particle are highlighted with a plus and a cross, respectively.

### 4 Method – the parallax effect

Several trails in Fig. 3 show a parallel shift between the green (NAC) and red (WAC) trail, like the large faint two trails crossing the image from top left to bottom right. Figure 7 presents an enlarged particle with a clear shift. It is included in the second image set taken on 24 July 2015 at about 15:25 UTC. Unfortunately, parts of the trails lie outside of the image. For this reason, only one part of the NAC trail is included as well as only part of the WAC trail. First, the beginning and end of the trail has to be found. For that purpose, a detection program was written.

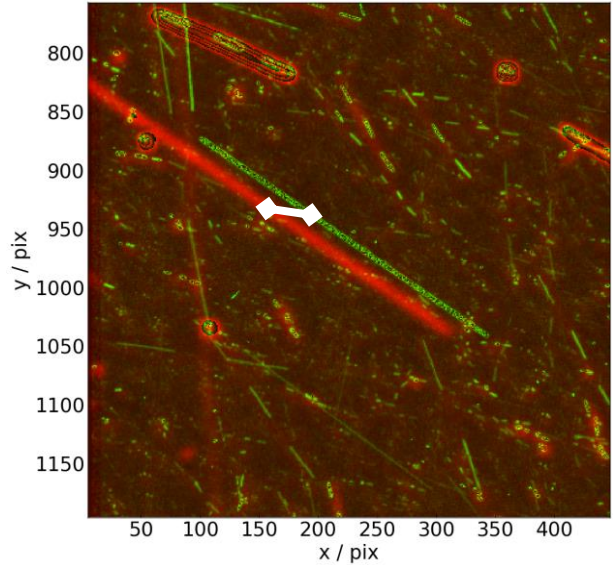

**Figure 7** Enlarged particle with a clear shift. It is included in the second image set taken on 24 July 2015 at about 15:25 UTC, which shows a parallel shift between the green higher resolution (NAC) and red lower resolution (WAC) trails. The white connecting line represents the length of the shift between the two particle trails. The pixel values presented are in the NAC resolution. The filenames of the utilized images are: NAC_2015-07-24T15.24.57.858Z_ID30_1397549000_F41, NAC_2015-07-24T15.25.21.725Z_ID30_1397549001_F22, NAC_2015-07-24T15.25.45.756Z_ID30_1397549002_F24, and WAC_2015-07-4T15.24.53.787Z_ID30_1397549500_F21.

It is a half-automatic program. The user has to identify a part of the original image including (ideally only one) trail that is meant to be investigated. With a canny edge detection algorithm (Canny, 1986) and subsequently a Hough transformation ((Hough, 1959) and (Duda and Hart, 1972)) the particle edges were found. A fit through the particle is computed using these edges. Afterwards, the result is further improved using the maxima of 1-dimensional Gaussians perpendicular to the line of flight. Next, the intensity profile of the particle's signal along its line of flight is found and an error function is fit to its flanks to compare the results with the theory presented in Section 3.2.1. The positions where the intensity has fallen to half its maximum are then declared to be the start respectively end point of the particle. To deal with the noise in the images, the minimal intensity value of the particle vicinity is subtracted from the data. Additionally, the error function fitted to the flanks provides a smoothed result that includes the noise left in the background as well as of the profile. The algorithm is further explained in Ott et al. (2016). With the start and end points of the particles' trails the mean shift between them can be computed. With this shift $a$, the distance between the cameras $d_{Cam}$, and the spatial resolution of the NAC of 0.0011 °/pixel the distance to the cameras can be computed as follows

$$d_{Particle} = \frac{\frac{d_{Cam}}{2}}{\tan\left(\frac{\alpha}{2}\right)} \approx \frac{d_{Cam}}{\alpha}, \qquad (19)$$

where $\alpha = 0.0011$ °/pixel $\cdot a$, which is very small for our calculations. We point out that $\alpha$ has to be inserted in degrees, if it is the argument of the tangents and in radians after the approximation. The geometry of this parallax determination is presented in Fig. 8.

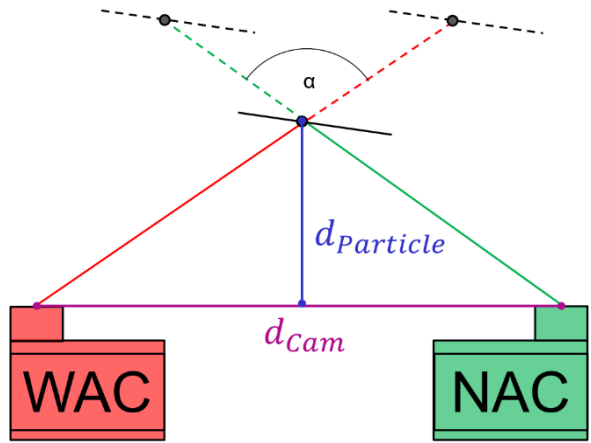

**Figure 8** The geometry of the parallax determination. From the point of view of the spacecraft the WAC is located left and the NAC right. The distance between the cameras is $d_{Cam} \approx 0.69$ m, the distance of the particle to the cameras is $d_{Particle}$. The shift between the trails determines the angle $\alpha$.

### 4.1 X-Y-Parallax

Figure 9 right shows a sketch of the front view of the Rosetta spacecraft. The geometry of the cameras is shown, the NAC on the left and the WAC on the right. The minimum distance between the centers of the cameras is drawn in (dashed line). The z-axis is directed to point outwards of the image plane. The distance components in all directions are taken from a Rosetta model from de Wilde (pers. comm., 2016):

$x_{dist} = 688.63$ mm  (20)
$y_{dist} = 83.79$ mm  (21)
$z_{dist} = 57.50$ mm  (22)

The distance in z-direction can be neglected due to the fact that the particles that show a parallax effect are at a distance of several kilometers from the spacecraft. Nonetheless, the distances in x-direction as well as in y-direction influence the parallax shift of the particle.

Consequently, the distance between the two cameras and the shift between the particles in the same direction has to be used. In other words, if one would consider the direct line between the two cameras as a line on a plane, the shift should be in the same plane. If the cameras were only shifted in the x-direction, the plane would be horizontal, compare Fig. 9 left. Hence, we would also see a shift in the horizontal direction. If the cameras were now also shifted in the y-direction, the connecting line would be tilted by an angle. This would lead to a tilted plane in which the shift lies.

Since the OSIRIS cameras are mounted in such a way that the image coordinate systems are aligned, their x-axes are parallel to each other. But, the plane on which the parallax shift occurs is tilted by an angle. Accordingly, the parallax shift will also be tilted by that angle. That is why the distance between the two trails of the same particle is a line between them, tilted by the same angle towards the horizontal plane. Compare Fig. 9 right that shows a shift between the trails in the resulting image caused by the parallax effect.

According to this, the distances to the cameras has to be determined with the absolute distance between the two cameras

$$d_{Cam} = \sqrt{83.79^2 + 688.63^2} \text{ mm } = 693.71 \text{ mm}, \qquad (23)$$

and the measured shift between the two trails along an angled connecting line, compare the dashed line in Fig. 9 right. To compute this line, the angle $\gamma$ has to be computed as

$$\gamma = \arctan\left(\frac{x_{dist}}{y_{dist}}\right) = 6.94°. \qquad (24)$$

Equation (19) yields the distance between the spacecraft and the particle with the minimal distance between the cameras $d_{cam}$ and the angular shift $\alpha$, where $\alpha = a \cdot 0.0011°$/pixel, with the measured shift in pixel $a$. In this case the shift $a_{tilted}$ is at an angle $\gamma$ to the horizontal shift $a_x$. This is the shift between the trails that will be used for the following calculations.

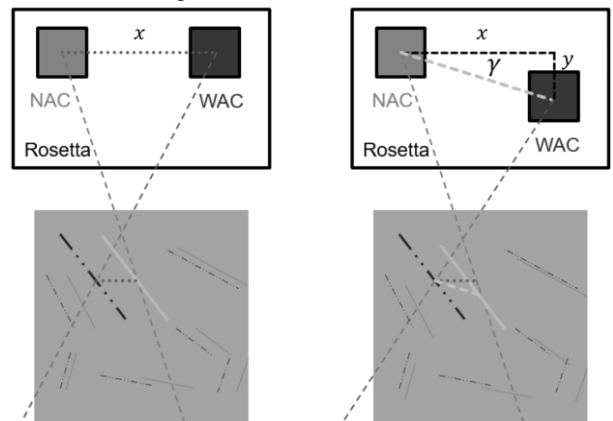

**Figure 9** Sketch of a shift between the trails in the resulting image caused by the parallax effect. Solid line: Trail as seen by NAC; dash-dot-dotted line: trail as seen by WAC; left: if the cameras are only shifted in x-direction. The dotted line shows the camera and particle shift in the same plane; right: if the cameras are shifted in x- and y-direction. The dotted line shows the shift if the cameras are shifted only in x-direction and the dashed line shows a camera shift in x- and in y-direction.

The images were taken at nearly the same time. This minimal deviation can be neglected for the dynamics of the particles. Nonetheless, it is needed to keep that in mind for the parallax determination. It is the reason why the tilted shift at the middle of the trails is used and not (and easier) the direct distance between the start positions of the NAC and WAC trails.

## 5 Results

The analysis explained in the previous Sections was done for all particles showing a shifted trail in the analyzed double camera data. In the following the investigation of one example particle will be presented.

### 5.1 Example

The trail presented in Fig. 7 results in a shift $a = 23.92$ pixel of the NAC (compare white line in Fig. 7) yielding an angular parallax angle of $\alpha = 0.026°$. Using this angle $\alpha$ and $d_{Cam}$ from Eq. (23), the distance of the dust particle to the spacecraft can be computed by Eq. (19) to be $d_{Particle} \approx 1511$ m.

### 5.2 Error analysis

For the distance determination an error analysis has been done. Error propagation yields Eq. (25). First, the different error values were estimated. The used interpolation is a source of error. Additionally, the alignment of the images is not perfect. We estimate the star matching accuracy to be approximately 0.5 pixel. Moreover, the determination of the beginning and end position of the particles is imprecise due to the faintness of most of the shifted trails. Here an error of about 0.5 pixel is assumed. As a result, the error of $\Delta a = 1$ pixel is assumed.

Additionally, for the distance between the cameras $d_{Cam}$, that has small uncertainties, an accuracy of up to 0.5 cm is assumed.

$$\Delta d_{Particle} \approx$$
$$\sqrt{\left(\left|\frac{1}{a \cdot 0.0011 \cdot \frac{\pi}{180}}\right| \cdot \Delta d_{cam}\right)^2 + \left(\left|\frac{-d_{Cam}}{0.0011 \cdot \frac{\pi}{180} \cdot a^2}\right| \cdot \Delta a\right)^2} \quad (25)$$

For the example presented in Fig. 7 this leads to
$$d_{Particle} = (1511 \pm 63) \text{ m}. \quad (26)$$
Other error estimations can also be considered.

### 5.3 Expected results

In the whole data set there are more than 250 promising particles showing a shifted trail. For those particles a distance analysis can be carried out using the method derived in this paper. Distances in the range of 500 m – 6000 m are expected.

For closer particles the trail will be out of focus. We found a few of such particles with a blurred signal making a shift determination between the trails quite uncertain. For particles that are further than 6 km from the spacecraft, the recorded shifts become too small and fall into the error margin. The question is: how accurately can we measure very small shifts? There is already an uncertainty due to the larger FOV of the WAC with the same pixel size as the NAC. One WAC pixel corresponds to six NAC pixels. For particles that are 6000 m away from the spacecraft the shift between the trails is about six pixels. Moreover, the trails have a width of about three pixels in the NAC and of about two pixels in the WAC (corresponding to 12 NAC pixels). Even with an extrapolation of the WAC signal to convert it to the NAC dimensions, shifts that are smaller than six pixels would yield signals too close to each other. Most probably, shifts smaller than three pixels would yield overlapping NAC and WAC trails. Additionally, the uncertainty increases with growing distances. For particles that are relatively close to the spacecraft, as the example presented above, the error is smaller than 5% of the value. However, for particles that are 6 km away from the spacecraft the error increases to more than 15 %. All in all, we assume that the minimal measurable trail distance is six pixels.

Additionally, to deriving a distance distribution for the whole data set, it is also possible to determine the velocity of the particles taking the spacecraft motion into account and assuming a radial motion away from the nucleus. Furthermore, their brightness can be determined, utilizing some assumptions. With this information and the exposure time, the size of the particles as well as a particle flux can be calculated. We expect the particles to not be too small but rather in the range of g to kg since we believe we can derive particles up to a distance of 6 km. Additionally, our data is comparable to the data used by Fulle et al. (2016) who found particles in a mass range of $10^{-4} - 10^2$ kg.

## 6. Conclusion

In this work it has been shown that it is possible to determine the distance of particles from Rosetta by simultaneous OSIRIS observations utilizing the parallax effect. To do so, the intensity profile of the recorded signal was analyzed and the positions of the start and end of the trail were computed. These positions are both found to be the point where the error function fit, to the intensity profile of the trail, has fallen to half its maximum, as shown with the theoretical particle.

One example particle was analyzed yielding a distance to the spacecraft of about 1.5 km. The expected results for other simultaneous observations are in the range of 500 m – 6000 m. For nearer particles the signal will become unfocused. For particles that are further away, the shifts will become too small to be detectable. The investigated particle is inside the expected distance range.

In the future the whole data set will be analyzed and a distance distribution of more than 250 promising particles is expected. In a second step, the velocity and brightness can be determined yielding a size as well as a mass distribution. Fulle et al. (2016) derived a mass distribution for the same time period, resulting in a distribution of masses in the range of $10^{-4} - 10^2$ kg. This is the mass range we expect for our particles, too.

In total, there are seven days of observations available resulting in 40 image sets including promising particles, corresponding to 138 NAC frames plus 40 WAC frames. The observation conditions change for the different days, in half the data even for each set of imaging sequences.

With the known start and end points of the particles and the known distance it will be possible to compute the velocity of the particles. We expect the spacecraft motion to not affect the apparent particle motion by a lot, however the orientation of the spacecraft with respect to the comet nucleus has to be taken into consideration. Again, the geometry changes partially even for different image sets and has to be investigated individually.

Analyzing all these data will yield the possibility to derive a particle density around the perihelion of 67P, when the data was collected. Again, a number density was published for a different method by Fulle et al. (2016). We expect comparable results.

**Acknowledgements**

OSIRIS was built by a consortium led by the Max-Planck-Institut für Sonnensystemforschung, Göttingen, Germany, in collaboration with CISAS, University of Padova, Italy, the Laboratoire d'Astrophysique de Marseille, France, the Instituto de Astrofísica de Andalucia, CSIC, Granada, Spain, the Science Support Office of the European Space Agency, Noordwijk, The Netherlands, the Instituto Nacional de Técnica Aeroespacial, Madrid, Spain, the Universidad Politéchnica de Madrid, Spain, the Department of Physics and Astronomy of Uppsala University, Sweden, and the Institut für Datentechnik und Kommunikationsnetze der Technischen Universität Braunschweig, Germany.

Rosetta is an ESA mission with contributions from its member states and NASA. The support of the national funding agencies of Germany (DLR), France (CNES), Italy (ASI), Spain (MEC), Sweden (SNSB), and the ESA Technical Directorate is gratefully acknowledged.

We thank the Rosetta Science Ground Segment at ESAC, the Rosetta Mission Operations Centre at ESOC and the Rosetta Project at ESTEC for their outstanding work enabling the science return of the Rosetta Mission.

We thank one of the referees for pointing out the simplification of the calculations presented in Eq. (14).

**References**

[1] Andrews, Special Functions of Mathematics for Engineers. SPIE Press; Auflage: 2 (1998).
[2] Canny, A Computational Approach to Edge Detection. IEEE Transactions Pattern Analysis and Machine Intelligence, 8(6), pp. 679-698 (1986).
[3] Davidsson, et al., Orbital elements of the material surrounding comet 67P/Churyumov-Gerasimenko. A&A, Volume 583, 16D (2015).
[4] de Wilde, pers. comm., Screenshot of Rosetta_29_03_2007_DW.exe (2016).
[5] Della Corte et al., GIADA: Its Status After the Rosetta Cruise Phase and On-Ground Activity in Support of the Encounter with Comet 67P/Churyumov-Gerasimenko, Journal of Astronomical Instrumentation, pp. Vol. 3, No. 1 (2014).
[6] Duda & Hart, Use of the Hough Transformation to Detect Lines and Curves in Pictures, Comm. ACM, Vol. 15, No. 1, pp. 11-15, (1972).
[7] Fraser et al., TRIPPy: Trailed Image Photometry in Python, The Astronomical Journal, Vol. 151, No 6 (2016).
[8] Fulle et al., Evolution of the dust size distribution of comet 67P/Churyumov-Gerasimenko from 2.2 AU to Perihelion, The Astrophysical Journal, Vol. 821, No 1 (2016).
[9] Hough, Machine Analysis of Bubble Chamber Pictures. Proc. Int. Conf. High Energy Accelerators and Instrumentation (1959).
[10] Keller et al., OSIRIS - The Scientific Camera System Onboard Rosetta, Space Science Reviews, Vol. 128, Issue 1, pp. 433-506 (2007).
[11] Magrin et al., Pre-hibernation performances of the OSIRIS cameras onboard the Rosetta spacecraft, Astronomy & Astrophysics, 574: A123 (2015).
[12] Ott et al., PaDe - The Particle Detection Programm. Proceedings of the IMC 2016 Egmond, NL (2016).
[13] Pleiades Astrophoto S.L. 1.8, PixInsight, Retrieved 15 06, 2016, from http://pixinsight.com/ (2016).
[14] Rotundi et al., Dust measurements in the coma of comet 67P/Churyumov-Gerasimenko inbound to the sun. Science, Vol. 347, Issue 6220 (2015).
[15] Tubiana et al., Scientific assessment of the quality of OSIRIS images, Astronomy & Astrophysics, Vol. 583, A46 (2015).
[16] Vereš et al., Improved Asteroid Astrometry and Photometry with Trail Fitting, PASP, 124, 1197 (2012).

**Affiliations**

[1]University of Oldenburg, Ammerländer Heerstraße 114, 26111 Oldenburg, Germany.
[2]ESA/ESTEC, 2201 AZ Noordwijk ZH, The Netherlands
[3]Max-Planck-Institut für Sonnensystemforschung, 37077 Göttingen, Germany
[4]Chair of Astronautics, TU Munich, Germany.
[5]Department of Physics and Astronomy, University of Padova, 35122 Padova, Italy.
[6]Laboratoire d'Astrophysique de Marseille, UMR 7326, CNRS & Aix Marseille Université, 13388 Marseille Cedex 13, France.
[7]Centro de Astrobiologia, CSIC-INTA, 28850 Madrid, Spain.
[8]International Space Science Institute, 3012 Bern, Switzerland.
[9]Department of Physics and Astronomy, Uppsala University, 75120 Uppsala, Sweden.
[10]PAS Space Research Center, 00716 Warszawa, Poland.
[11]Department of Astronomy, University of Maryland, College Park, MD 20742, USA.
[12]Akademie der Wissenschaften zu Göttingen, 37077 Göttingen, Germany.

[13]LESIA–Observatoire de Paris, CNRS, Université Pierre et Marie Curie, Université Paris Diderot, 92195 Meudon, France.

[14]LATMOS, CNRS/UVSQ/IPSL, 78280 Guyancourt, France.

[15]Dipartimento di Fisica e Astronomia "G.Galilei", University of Padova, Vicolo dell'osservatorio 3, 35122 Padova, Italy.

[16]INAF, Osservatorio Astronomico di Padova, 35122 Padova, Italy.

[17]CNR-IFN UOS Padova LUXOR, 35131 Padova, Italy.

[18]Department of Industrial Engineering, University of Padova, 35131 Padova, Italy.

[19]INAF - Osservatorio Astronomico di Trieste, 34014 Trieste, Italy.

[20]Aix Marseille Université, CNRS, Laboratoire d'Astrophysique de Marseille, UMR 7326, 13388 Marseille, France.

[21]Instituto de Astrofisica de Andalucia (CSIC), c/ Glorieta de la AstronomÌa s/n, 18008 Granada, Spain.

[22]Deutsches Zentrum für Luft- und Raumfahrt, Institut für Planetenforschung, 12489 Berlin, Germany.

[23]Graduate Institute of Astronomy, National Central University, Chung-Li 32054, Taiwan.

[24]Space Science Institute, Macau University of Science and Technology, Macao, China.

[25]Institut für Geophysik und extraterrestrische Physik, Technische Universität Braunschweig, 38106 Braunschweig, Germany.

[26]Operations Department, European Space Astronomy Centre/ESA, 28691Villanueva de la Canada, Madrid, Spain.

[27]Department of Information Engineering, University of Padova, 35131Padova, Italy.

[28]Physikalisches Institut, University of Bern, 3012 Bern, Switzerland.

[29]Center for Space and Habitability, University of Bern, 3012 Bern, Switzerland.